\documentclass[twocolumn]{article}
\usepackage{graphicx}

\begin{document}

\thispagestyle{empty}

\begin{center}
\begin{minipage}[t]{0.48\textwidth}
\scriptsize
\begin{tabular}{|c|}
\hline
~~~~~~~~~~~~~~~~~~~~~~~\includegraphics[scale=1.0]{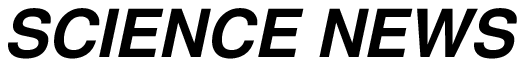}~~~~~~~~~~~~~~~~~~~~~~~\\
\hline
\end{tabular}
\end{minipage}
\end{center}

{\large \bf Are Passive Spiral Galaxies truly ``passive'' and ``spiral''? : a Near-Infrared perspective}

\vspace*{1mm}

{\it Chisato Yamauchi}$^{\mbox{\scriptsize \it 1,2}}$ {\it and}
{\it Tomotsugu Goto}$^{\mbox{\scriptsize \it 3}}$

\vspace*{1mm}

\begin{minipage}[t]{0.48\textwidth}
\scriptsize
$^{\mbox{\tiny \it 1}}${\it Department of Physics and Astrophysics, Nagoya University, Chikusa-ku, Nagoya 464-8602, Japan}\\
$^{\mbox{\tiny \it 2}}${\it National Astronomical Observatory, 2-21-1 Osawa, Mitaka, Tokyo 181-8588, Japan}\\
$^{\mbox{\tiny \it 3}}${\it Department of Physics and Astronomy, The Johns Hopkins University, 3400 North Charles Street, Baltimore, MD 21218-2686, USA}\\
\end{minipage}

Passive spiral galaxies - unusual galaxies with spiral morphologies but without any sign of on-going star formation - have recently been shown to exist preferentially in cluster infalling regions. This discovery directly connects passive spiral galaxies to cluster galaxy evolution studies, such as the Butcher-Oemler effect or the morphology-density relation. Thus, detailed study of passive spiral galaxies could potentially yield new insight on the underlying physical mechanisms governing cluster galaxy evolution.

\parindent 13pt

However, in previous work, passive spiral galaxies were selected from low-resolution optical images with $\sim$1.5 arcsec seeing. Passive spirals could therefore be mis-identified as S0 galaxies, or as dusty-starburst galaxies which are not passive at all. To address this issue we performed deep, high-resolution, near-infrared imaging of 32 passive spiral galaxies with UKIRT.

We selected our target galaxies from 73 passive spiral galaxies
presented by Goto et al. (2003). None of the 73 galaxies have any
emission in [OII] or H$\alpha$ ($<$1$\sigma$ in equivalent width)
though all have a disc-like morphology. Of the 73 passive spiral
galaxies, the 32 targets accessible during our run in September 2003
were observed in the K band using the UKIRT Fast Track Imager
(UFTI). Data were taken during
periods of good atmospheric transparency
and with excellent seeing of $\sim$ 0.5 arcsec. 
In Figure 1, we show 
{\it K}-band ~images {\scriptsize ~~}of {\scriptsize ~~}16 {\scriptsize ~~}of {\scriptsize ~~}the {\scriptsize ~~}32 {\scriptsize ~~}passive {\scriptsize ~~}spiral {\scriptsize ~~}galaxies. 

\begin{figure}[!h] 
\includegraphics[scale=0.44]{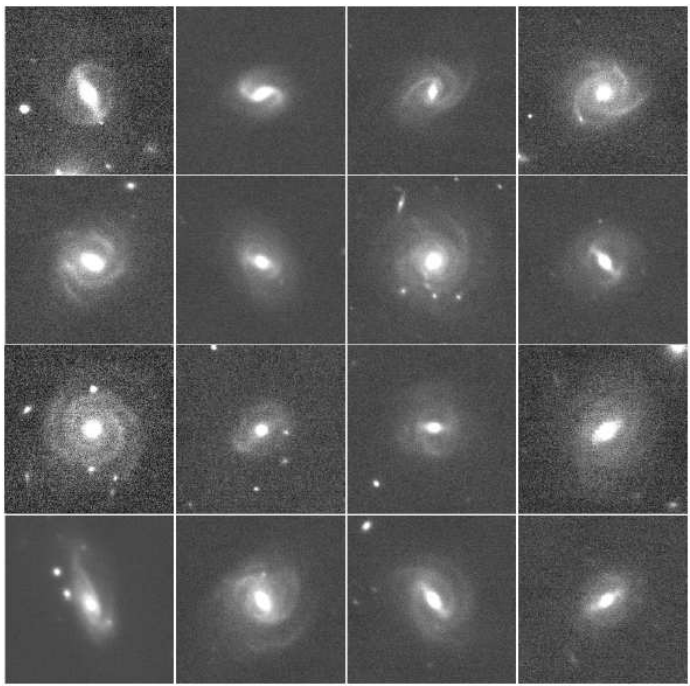}
\vspace*{-3mm}
\caption{
UKIRT {\it K} band images of passive spiral galaxies. Each image is 35x35 arcsec in size.}
\end{figure}

\parindent 0pt

Although
selected in poorer conditions, the deep and high resolution
imaging
capability of UKIRT clearly shows the discs and spiral arm
structures. Thus, passive spiral galaxies are not S0s, but truly are
spiral galaxies.

\parindent 13pt

We used the restframe optical-infrared ({\it r}-{\it K}) colour distribution for the
observed passive spiral galaxies to investigate whether they are dusty
starburst galaxies or truly passive galaxies. Since the K band is less
affected by dust extinction than the r band, dusty starburst galaxies
are known to have redder colours in {\it r}-{\it K} by $\sim$1 mag (Smail et
al. 1999). Figure 2 plots {\it g}-{\it i} colour against {\it r}-{\it K} colour. Optical
photometry ({\it g}, {\it r}, and {\it i}) is from the SDSS. The black circles are for
passive spiral galaxies observed with UKIRT. The red squares are for
early-type galaxies in the control sample. For a reference, we plot the
distribution of all galaxies in the volume limited sample with {\it K}
magnitudes measured with the Two Micron All Sky Survey (2MASS; Jarrett
et al. 2000) as the contour.

Interestingly, compared with all galaxies (the contour), passive spiral
galaxies (circles) are not redder in {\it r}-{\it K} colour. Indeed, the
{\it r}-{\it K} colours
of the passive spiral galaxies are indistinguishable from the early-type
galaxies (squares). These results support the truly passive nature of
these galaxies, since dusty starburst galaxies should have {\it r}-{\it K} colours
redder by 1 magnitude than normal galaxies.

Thus, our results support the truly ``passive'' and ``spiral'' nature of
these galaxies. It is very likely that passive spiral galaxies are
indeed transition objects currently undergoing cluster galaxy
evolution. Further studies of passive spiral galaxies will reveal the
physical mechanisms governing cluster galaxy evolution.

\parindent 0pt

\begin{minipage}[t]{0.48\textwidth}
\scriptsize
~\\
{\bf References}\\
Goto T., Okamura S., Sekiguchi, M., et al. 2003, PASJ, 55, 757\\
Jarrett T. H., Chester T., Cutri R., Schneider S., Skrutskie M., Huchra J. P., 2000, AJ, 119, 2498\\
Smail I., Morrison G., Gray M. E., Owen F. N., Ivison R. J., Kneib\\
J.-P., Ellis R. S., 1999, ApJ, 525, 609\\
\end{minipage}

\vspace*{-8mm}

\begin{figure}[!b] 
~~~\includegraphics[scale=0.42]{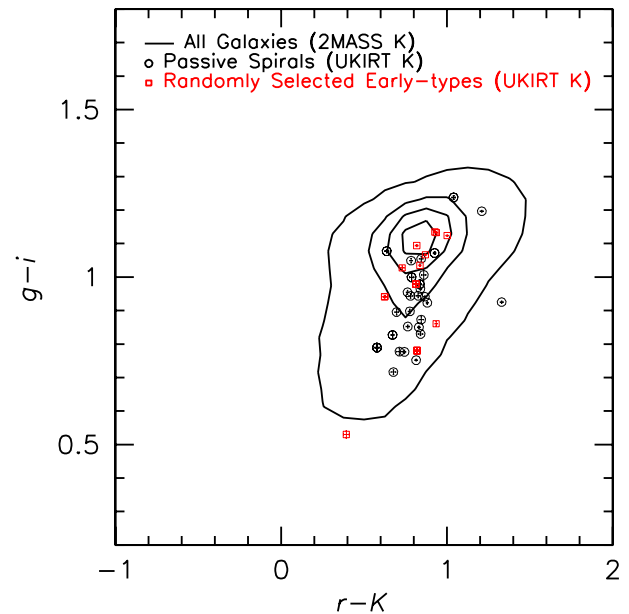}
\vspace*{-2mm}
\caption{
Restframe {\it g}-{\it i} vs. {\it r}-{\it K} two-colour diagram. The
 circles are for passive spirals. The squares are for the early-type
 galaxies in the control sample. The contours represent all galaxies in
 the volume limited sample with 2MASS {\it K} magnitudes.}
\end{figure}

\end{document}